# Leakage-Aware Reallocation for Periodic Real-Time Tasks on Multicore Processors

Hongtao Huang, Feng Xia, Jijie Wang, Siyu Lei, and Guowei Wu
School of Software
Dalian University of Technology
Dalian 116620, China
f.xia@ieee.org

**Abstract**

*It is an increasingly important issue to reduce the energy consumption of computing systems. In this paper, we consider partition based energy-aware scheduling of periodic real-time tasks on multicore processors. The scheduling exploits dynamic voltage scaling (DVS) and core sleep scheduling to reduce both dynamic and leakage energy consumption. If the overhead of core state switching is non-negligible, however, the performance of this scheduling strategy in terms of energy efficiency might degrade. To achieve further energy saving, we extend the static task scheduling with run-time task reallocation. The basic idea is to aggregate idle time among cores so that as many cores as possible could be put into sleep in a way that the overall energy consumption is reduced. Simulation results show that the proposed approach results in up to 20% energy saving over traditional leakage-aware DVS.*

## 1. Introduction

Processor slowdown and shutdown are two popular techniques to reduce power consumption in real-time embedded systems. In the realm of processor slowdown, dynamic voltage scaling (DVS) has proved to be an energy efficient technique, which reduces dynamic power by scaling down processor voltage [8]. While DVS reduces the dynamic power, it increases leakage energy consumption (caused by leakage currents). This is because the lowered processor speed lengthens the execution time of tasks. On the other hand, studies have explored the use of processor shutdown to reduce leakage energy, i.e. turning an idle processor/core into sleep state when the computational workload is light [5, 7]. To make DVS leakage-aware, the concept of critical speed has been introduced, which tries to balance the use of processor slowdown and shutdown by posing a restriction on the minimum speed that pure DVS scales [5]. When the workload is low, a processor usually runs at a speed higher than pure DVS instructs. In this case, the processor idle time exists, and it is possible to exploit this idle time to achieve further energy saving by turning the processor into sleep state.

In real-time systems, energy efficient techniques also need to make sure the lowered computational performance does not violate the schedulability of the tasks. As a result, energy-aware scheduling is introduced to reduce energy consumption while still completing tasks in time [9, 1]. In the past decades, significant efforts have been made on energy-aware scheduling that considers leakage energy consumption[11, 3, 5, 2, 4]. However, most of the studies address uniprocessor or multiprocessor, and very few address multicore processors.

Multicore processors (or chip multiprocessors), which integrate multiple processing units (i.e. cores) into a single chip, have emerged as a remedy to reduce power consumption caused by the increasing processing capabilities of processors. However, the leakage energy consumption of multicore processors has risen due to the vastly increased number of circuits [9]. Moreover, some multicore processors cannot independently adjust the clock frequency and voltage of cores. Blindly applying existing energy-aware scheduling for multiprocessors to these multicore processors may possibly incur extra energy consumption [9].

In this work, we address leakage-aware scheduling for periodic real-time tasks on multicore processors, where cores can enter sleep state at any idle time. Studies have shown that DVS with critical speed (leakage-aware DVS) may incur more energy consumption if the overhead of core state switching is non-negligible [5]. We will illustrate that on multicore platforms, a balanced task-core assignment may further impair the energy efficiency of leakage-aware

DVS. We then present a task reallocation algorithm to improve the energy efficiency of leakage-aware DVS by analyzing run-time idle intervals and making aggressive task allocation decision. We also evaluate the energy efficiency of the task reallocation strategy with various simulation settings.

The rest of the paper is organized as follows. Section 2 summarizes prior research on leakage-aware scheduling. Section 3 provides task model, power model, and concepts used in our work. Section 4 explains the motivation for reallocation. Section 5 presents the reallocation algorithm. Section 6 provides an evaluation based on simulation results. Section 7 presents our concluding remarks.

## 2. Related Work

For uniprocessors, many studies were made on the leakage-aware scheduling for periodic real-time tasks in DVS system [5, 11, 3]. Jejurikar et al. raised the concept of *critical speed* for DVS in [5], where they adopted *energy per cycle* as a better metric for energy consumption. Based on the critical speed, both fixed-priority (e.g. RM) and dynamic-priority (e.g. EDF) scheduling on uniprocessors were extended to be leakage aware. For instance, Zhu et al. [11] used the feedback actual execution time to adjust the time to turn processor into sleep state in idle time. For RM scheduling, Chen et al.[3] devised a scheduler to aggregate idle intervals by revising the arrival time of tasks based on on-line simulation of RM scheduling.

For multiprocessors, leakage-aware scheduling of periodic real-time tasks mainly focuses on the task-processor assignment. Partition approach is widely adopted in the task assignment, where a task set is statically partitioned into several sub-task sets and assigned to processors [2, 4]. After the partition, energy-aware task scheduling for uniprocessors is applied to each processor. For example, Chen et al. [2] proposed two partition algorithms based on the largest-task-first (LTF) heuristic, with shutdown overhead addressed.

Recently, researchers in leakage-aware scheduling have started to address multicore processors [10, 9]. Off-line approaches were considered by Chen et al. in [10] for frame-based periodic tasks, where they decided the processor speed as well as time to turn core into sleep state based on LTF partition. For on-line approach, Seo et al. [9] considered scheduling of periodic real-time tasks through run-time task reallocation. When a task completes prior to its worst case estimation, the proposed scheduler reallocates all the tasks among cores to balance workloads. It also dynamically scales the active core number through task reallocation with the knowledge of the optimal core number.

In this work, we extend the partition based scheduling on multiprocessors/multicore processors through task reallocation, which allows a task to be shifted to another core during run-time. A similar approach is also taken in [9]. However, our approach only reallocates a task when its new instance (job) arrives, but rejects further reallocation of the job during its execution time. Also, unlike previous work on multicore processor [10, 9], we take into account the overhead of core state switching.

## 3. Preliminaries

In this section, we first give the task and power model, and then briefly describe the concept of global DVS, task reallocation and critical speed.

### 3.1. Task Model

We consider a task set $T = \{\tau_1, \tau_2, \tau_3, ..., \tau_n\}$ of $n$ periodic, soft real-time tasks. Each task $\tau_i$ is modeled as $\tau_i = (P_i, W_i)$, where $P_i$ is the period, and $W_i$ is the worst case execution time (WCET) when task is running with the maximum speed. The relative deadline of a task is equal to its period in our model. Tasks are assumed independent of each other. A job is the periodically released instance of a task. The $j_{th}$ job of $\tau_i$ is with arrival time $a_{ij} = (j-1) \cdot P_i$ and deadline $d_{ij} = j \cdot P_i$.

The utilization $u_i$ of $\tau_i$ is given by $u_i = W_i/P_i$. Since most of the task finishes earlier than its worst case estimation, the dynamic utilization $u'_i$ of $\tau_i$ is updated upon its arrival and completion time:

$$u'_i = \begin{cases} W_i/P_i & \tau_i \text{ is not finished} \quad (1) \\ cc_i/P_i & \tau_i \text{ is finished} \quad (2) \end{cases}$$

where $cc_i$ is the actual execution time for the nearest invocation.

Task set $T$ is partitioned among a set of $m$ identical cores $C = \{C_1, C_2, ..., C_m\}$, and $T_j$ denotes the task set allocated to $C_j$. Tasks are preemptive and are scheduled by the Earliest Deadline First (EDF) scheduling policy on its partitioned core. The utilization of the task set $T_j$ on $C_j$ is defined by $U(C_j) = \sum_{\forall \tau_i \in T_j} u_i$. The dynamic utilization $U'(C_j)$ of the task set $T_j$ is given by $U'(C_j) = \sum_{\forall \tau_i \in T_j} u'_i$. We define $U_{tot}$ as the total utilization of the task set. We only consider task set with $U_{tot} <= m$, where $m$ is the number of cores.

### 3.2. Power Model

We adopt the power model based on Martin et. al [6]. The power consumed in a CMOS-based processor consists of two portions: dynamic power and static power. The dynamic power, $P_{AC}$ is given by:

$$P_{AC} = C_{eff}V_{dd}^2 f$$

where $C_{eff}$ is the average switched capacitance per cycle, $V_{dd}$ is the supply voltage, and $f$ is the clock frequency.

Static power $P_{DC}$ is caused mainly by the sub-threshold leakage current $I_{sub}$ and reverse bias junction current $I_j$, and is given by:

$$P_{DC} = L_g(V_{dd}I_{subn} + |V_{bs}|I_j)$$

where $V_{bs}$ is the body bias voltage, $L_g$ is the number of component in the circuit. The sub-threshold leakage current is given by:

$$I_{subn} = K_3 e^{K_4 V_{dd}} e^{K_5 V_{bs}}$$

where $K_3$, $K_4$ and $K_5$ are all constant parameters.

Cores are homogenous and share the same clock frequency. We consider continuous frequency, which is scaled between $f_{min}$ and $f_{max}$ with the change of supply voltage $V_{dd}$:

$$f = \frac{(V_{dd} - V_{th})^\epsilon}{L_d K_6}$$

$$V_{th} = V_{th_1} - K_1 V_{dd} - K_2 V_{bs}$$

where $V_{th}$ is the threshold voltage, and $V_{th_1}$, $L_d$, $K_1$, $K_2$, $K_6$ are technological constants. The overhead of voltage transition is neglected. The constants for 70nm technology can be found in [9], which are also adopted by previous work on leakage-aware scheduling [11, 5].

A core has two states: active state and sleep state. In active state, each core consumes both dynamic and static power. In sleep state, only static power will be consumed, and it is assumed negligible in this work [9, 11].

We consider *switching overhead Esw* when a core switches from sleep state to active state, as in [11, 5, 2]. The switching overhead is caused by several factors, such as cold start misses in cache among other resource refresh overheads [5].

### 3.3. Global DVS

The DVS scales down the frequency of the processor by scaling down the supply voltage. We take advantage of runtime DVS [8] in this work. The runtime DVS adjusts the frequency (processor speed) according to the dynamic utilization of tasks. To make our work complementary to DVS on uniprocessor platforms, we apply the DVS to the core with the task set of the highest dynamic utilization, and set the global voltage and frequency accordingly. It makes sure that the schedulability of the tasks is not violated as long as the maximum utilization of tasks assigned to cores is below 100% [9]. Global frequency (or global speed) is used to refer to the frequency obtained from using global DVS.

### 3.4. Task Reallocation

We take the partition approach in task scheduling by assigning each core a task queue. Though any partition approach on multiprocessors can be taken, we adopt the *worst fit* heuristic [1, 9], because it offers better timeliness/energy performance than other heuristic. One detailed implementation, the LTF scheduling, can be found in [2, 10]. The LTF partition sorts tasks according to static utilization in a non-increasing order and assigns each task to the core with the lowest utilization.

Unlike traditional partition scheduling, we consider runtime reallocation of tasks. For each task, we allow its jobs to be executed on different cores, but forbid any job to be executed on more than one core. In other words, the reallocation of a task can only be made when its new job arrives.

### 3.5. Critical Speed

The critical speed $f_{cri}$ can be deduced by solving $d(P_{AC} + P_{DC})/df = 0$ [5, 2]. For the 70$nm$ technology, the critical speed is calculated as 0.4 the maximum processor speed. We define the *critical scale factor* as the critical speed divided by the maximum processor speed.

Critical speed minimizes total energy consumption within a given period of time when leakage energy is considered [2]. The proof is based on the assumption that processor consumes no energy when it is idle. When the *switching overhead* is negligible, the assumption is achieved by turning idle cores into sleep state. With the knowledge of critical speed, leakage-aware DVS sets the scaled speed $f = min(max(f_{cri}, f'), f_{max})$, where $f'$ is the speed given by pure DVS.

When the *switching overhead* is non-negligible, DVS with critical speed is not always more energy efficient than that without critical speed. Idle intervals may be too short to generate enough energy saving. If the energy saved by putting an idle processor into sleep state is less than the switching overhead, processor shutdown increases energy consumption adversely. In this case, it will be more energy efficient to exploit idle time through speed slowdown at the very beginning. Previous studies have devised the *sleep threshold* as a compromise [11, 5, 3]. Let $T_{th}$ be $Esw/P_{idle}$, where $P_{idle}$ is the power consumption of processor by running at the minimum speed. If the idle time is longer than $T_{th}$, the processor can be put into sleep state. Otherwise, the processor runs at the lowest processor speed available. *Shutdown penalty* is used to refer to the additional energy consumed during processor idle time, when the processor with leakage-aware DVS cannot be turned into sleep state.

## 4. Motivation

In this section, we discuss the limitation of the balanced partition when it is incorporated with leakage-aware DVS. We then present the reasons for our consideration of task reallocation, and give a motivational example.

### 4.1. Limitation of Balanced Partition

When the task set is with high utilization, a balanced partition is preferable [5]. For multicore processors, a balanced partition lowers the global frequency of DVS, and thus reduces energy consumption [9]. However, when the maximum task utilization on cores is lower than the critical scale factor, a balanced partition is not most energy efficient, because it impairs the energy efficiency of leakage-aware DVS. Leakage-aware DVS instructs the processor to run at a speed no slower than the critical speed, and takes advantage of the idle time for core sleep. While the raised global speed increases total idle time, balancing the workload leads to an even distribution of idle time among cores and shortens the idle time each core. The shorter the idle time, the more likely that idle intervals will be shorter than the sleep threshold, and the more often a core will stay active when it is idle. Moreover, for multicore processors that cannot set the frequency of a core individually, it needs to apply the global speed to all cores. Therefore, when a core is idle, it might run at a speed higher than the minimum speed, which further increases energy consumption.

### 4.2. Requirements of Reallocation

As described in 4.1, balanced partition reduces the idle time on each core for light workload. To extend idle time, one may consider static approach. One possible method would be to modify the partition scheme to let it take unbalanced partitioning strategy for task set with low utilization, as in [2]. However, even with the increased total idle time per core, it is still not guaranteed that idle interval is longer than the sleep threshold because idle time can be scattered among cores as small pieces of fragments. Besides, a task's job may finish earlier than its WCET. Therefore, even if the static utilization of tasks assigned to a core is above the critical scale factor, dynamic utilization could still fall below the critical scale factor during run-time. If the slack time due to early task completion is shorter than the sleep threshold, it cannot be exploited for core sleep either.

To dynamically reduce the shutdown penalty of leakage-aware DVS, two issues need to be addressed: 1) minimize the times of state switching in order to reduce total switching overhead, and 2) minimize the times of cores failing to turn into sleep state. Merging the fragmentary idle intervals helps achieve these two goals, because it not only reduces the number of idle intervals but also lengthens the majority of idle intervals to be longer than the sleep threshold whenever possible.

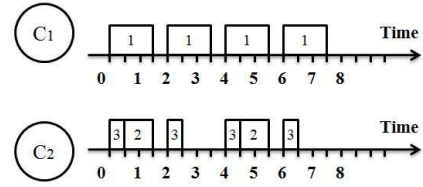

**Figure 1. Leakage-aware DVS without task reallocation**

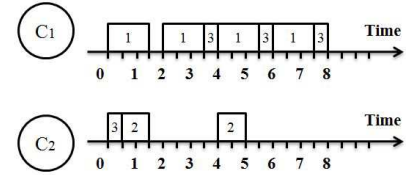

**Figure 2. Leakage-aware DVS with task reallocation**

Task reallocation can be used to merge the idle intervals. The basic idea is that if a core has several fragmentary idle intervals shorter than the sleep threshold, we can make use of these idle intervals for energy saving either by reducing the workload of the core or by adding more workload. For example, when a task comes between two idle intervals on a core, if we shift the task away to another core, we can then merge two fragment idle intervals. On the other hand, for the other idle cores that waste energy running at a speed higher than the minimum speed, the shifted task can run there and utilize the idle time.

### 4.3. Motivational Example

Consider a task set $T$ with three tasks, where $T = \{\tau_1(2, 0.6), \tau_2(4, 0.4), \tau_3(2, 0.2)\}$. To simplify the run-time decision of DVS, the acutal execution time for each task is set equal to the WCET. Suppose the critical speed is 40% the maximum speed, and the sleep threshold $T_{th}$ is 2 (units of time). Figure 1 shows the task execution after the LTF partition when DVS is performed. According to DVS, the speed could be reduced to 0.3 the maximum speed. With the consideration of leakage energy, we scale the speed up to the critical speed. Note that the scaled up speed leaves idle time for both cores. For core 1, the idle interval is 0.5, which is shorter than the $T_{th}$, therefore core 1 cannot turn into sleep state during idle time. For core 2, there are two types of idle intervals, neither of which is longer than $T_{th}$.

Figure 2 shows tasks execution after the reallocation, with DVS performed. The reallocation shifts task 3 to core 1 after its second invocation. Since the dynamic utilization of tasks after the reallocation is 0.4 for core 1 and 0.1 for core 2, the reallocation does not increase the global speed, and therefore does not increase dynamic power consumption. Furthermore, idle intervals on core 2 are merged and extended, making core 2 able to turn into sleep state during idle time to save energy.

## 5. The Proposed Algorithm

---

**Algorithm 1** Leakage-aware task reallocation

- $C_{home(\tau_i)}$: the core task $\tau_i$ is assigned to in the last allocation.
- $Queue(C_{home(\tau_i)})$: all tasks allocated to core $C_{home(\tau_i)}$ at current time
- $load(C_{home(\tau_i)})$: the accumulated cycles of tasks running on core $C_{home(\tau_i)}$. $load(C_{home(\tau_i)}) = \sum(W(\tau_j) \cdot f_{max})$ for each arrived task $\tau_j$ in $Queue(C_{home(\tau_i)})$ at time $t$.
- $C_{dest}$: the core to which $\tau_i$ will be assigned

1: upon_task_release($\tau_i$):
   // A job of $\tau_i$ arrives at time $t$
2: $dt \Leftarrow min(P_j \cdot \lfloor t/P_j \rfloor + P_j) - t - load[C_{home(\tau_i)}]/f_{cri}$, $\tau_j \in Queue(C_{home(\tau_i)})$
3: $C_{dest} \Leftarrow null$
4: **if** $dt + W(\tau_i) >= T_{th}$ **then**
5:     $C_{dest} \Leftarrow$ select_core($\tau_i$)
6: **end if**
7: **if** $C_{dest}\; != null$ **then**
8:     add $C_{home(\tau_i)}$ to $S$
9:     reallocate($\tau_i$,$C_{dest}$)
10: **else if** $C_{home(\tau_i)} \in S$ **then**
11:     remove $C_{home(\tau_i)}$ from $S$
12: **end if**

13: select_core($\tau_i$):
14: $C_{min} \Leftarrow$ core $C_i$ with minium $U'(C_i)$ **such that** $C_i \in S, U(C_i) + u(\tau_i) <= 1$, and $C_i! = C_{home(\tau_i)}$
15: **if** $U'(C_{min}) + u(\tau_i) <= f_{cri}/f_{max}$ **then**
16:     **return** $C_{min}$
17: **end if**
18: **return** $null$

19: reallocate($\tau_i$,$C_{dest}$):
20: remove $\tau_i$ from Queue($C_{home(\tau_i)}$)
21: insert $\tau_i$ into Queue($C_{dest}$)

---

The reallocation algorithm is invoked upon the arrival of a job. When a task's job arrives, we have to decide whether to shift the task to another core or let the task remain executing on the original core. We first calculate the minimum idle interval if the task is not shifted. Suppose at time $t$, a job of task $\tau_i$ arrives at core $C_{home(\tau_i)}$ and let $load(C_{home(\tau_i)})$ be the total cycle of the uncompleted jobs (including the newly arrived job) on core $C_{home(\tau_i)}$. $load(C_{home(\tau_i)})/f_{cri}$ is the maximum time to complete all the arrived jobs on $C_{home(\tau_i)}$. $Queue(C_{home(\tau_i)})$ is all tasks allocated to core $C_{home(\tau_i)}$ at $t$. The minimum idle interval $dt$ is the time interval between the time core $C_{home(\tau_i)}$ finishes all its arrived job and the arrival time of the next coming job on core $C_{home(\tau_i)}$, i.e. $dt = min(P_j \cdot \lfloor t/P_j \rfloor + P_j) - t - load(C_{home(\tau_i)})/f_{cri}, \tau_j \in C_{home(\tau_i)}$.

Then, we consider task reallocation. A greedy strategy is taken by first checking whether it is possible to reallocate the task to other cores. If $dt + W(\tau_i) >= T_{th}$, search is performed to find the destination core the task will shift to, because it is guaranteed that the idle interval is at least equal to the sleep threshold after shifting task $\tau_i$ to another core. The task will remain on its original core only if we fail to find the destination core or the idle time is too short. Such a strategy extends the idle interval when there is already idle time prior to the arrival of the current job.

To decide the core to reallocate the task to, we derive a heuristic to maintain a set $S$ of candidate cores. If a core fails to shift its latest job to other cores, we add the core into $S$, otherwise the core is removed from $S$ (if the core is in $S$). A task can be added to a core if it obeys the following rules:

(1) it does not make the utilization of the destination core above the critical scale factor; and

(2) it does not violate the schedulability of EDF [8].

Rule 1 is set to make sure that the global speed does not increase after the task reallocation. Among all cores in $S$, the core with the lowest dynamic utilization and consistent with the above rules will be chosen as the final destination core. The reason for only allocating a task onto cores in $S$ is that it makes sure a core that has shifted tasks away will not accept tasks from other cores before it turns into sleep state.

The pseudo-code for the leakage-aware task reallocation algorithm is shown in Algorithm 1. After the reallocation, EDF is applied according to the updated task assignment. We assume that there is a power manager that turns the core with idle time longer than the sleep threshold into sleep state and switches the sleeping core back to active when there is job ready to run on the core.

# 6. Simulations

## 6.1. Setup Overview

To evaluate the proposed reallocation algorithm, we developed a simulator based on the task and power model given in Section 3. We chose the technological constants that were also used by previous work [9, 11, 5]. The critical speed is 0.4 the maximum processor speed.

The simulator used randomly generated task sets, each containing up to 20 tasks. Tasks were assigned random periods within the range [10ms, 100ms]. This range is also used by others [11, 5, 2]. The total utilization was within (0,m], where $m$ is core number. Early completion of tasks is considered, with the ratio of actual execution time to WCET ($cc_i/W_i$) randomly generated at each invocation of tasks. The expectation of ($cc_i/W_i$) is within (0, 1].

Four different simulation settings were considered to explore the effect of the average total utilization ($U = U_{tot}/m$), switching overhead ($Esw$), the number of cores ($m$), and early completion of tasks ($cc$/WCET) on leakage-aware reallocation. Each setting was run 100 times independently.

Three algorithms were implemented in the simulations:

(1) Pure DVS: it uses LTF partition initially, and does not consider the critical speed in DVS;

(2) Leakage-aware DVS (LA-DVS): it takes LTF partition initially, and considers the critical speed in DVS, without run-time task reallocation; and

(3) Our approach denoted LA-Reallocation.

## 6.2. Results and Analysis

Representative results are given in Figures 3, 4, 5, and 6, where energy consumption is normalized to that of LA-DVS.

Figure 3 shows the simulation results by varying the average total utilization $U$. When $U$ is extremely low, LA-Reallocation does not save more energy than LA-DVS, because most of the idle intervals are longer than the sleep threshold and LA-DVS successfully exploits the idle time for core sleep. When $U$ rises to around the critical scale factor (i.e. 0.4), LA-Reallocation outperforms LA-DVS, with 20% peak energy saving. The reason is that, with the increased average total utilization, the percentage of the idle intervals shorter than the sleep threshold also increases, which makes LA-DVS fail to turn idle cores into sleep state. When $U$ rises to above 0.8, there is no energy saving of LA-Reallocation, because the dynamic utilization of tasks assigned to a core usually remains above the critical scale factor and leaves little room for task reallocation.

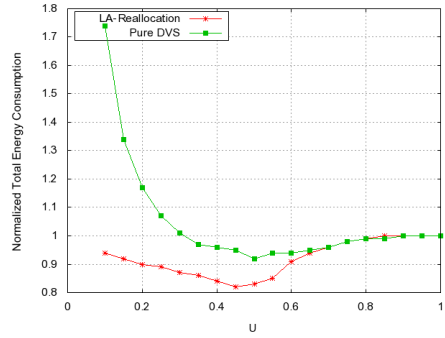

**Figure 3. Simulation result when** $Esw = 0.5mJ$**,** $m = 2$**,** $cc/$**WCET**$= 0.5$**, and** $U$ **ranges from** $0.1$ **to** $1$

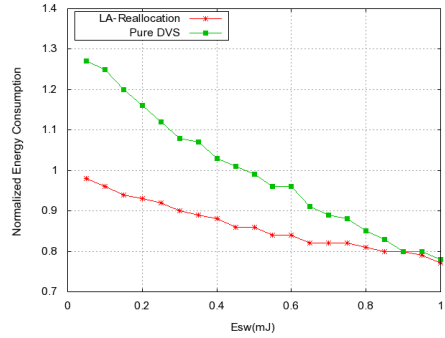

**Figure 4. Simulation result when** $m = 2$**,** $U = 0.3$**,** $cc/$**WECT**$= 0.5$**, and** $Esw$ **ranges from** $0$ **to** $1mJ$

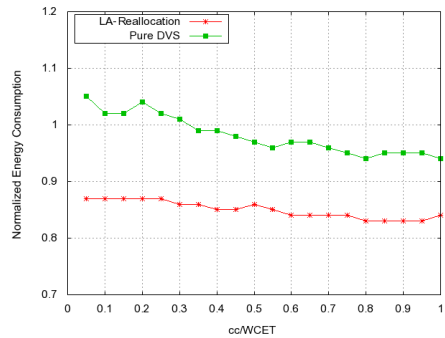

**Figure 5. Simulation result when** $Esw = 0.5mJ$**,** $m = 2$**,** $U = 0.3$**, and** $cc/$**WCET ranges from** $0.05$ **to** $1$

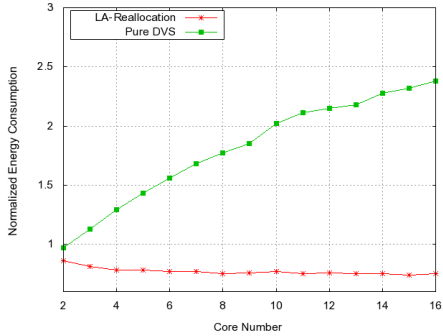

**Figure 6. Simulation result when** $Esw = 0.5mJ$**,** $cc/$**WECT**$= 0.5$**, and** $U = 0.3$**, core number ranges from** $2$ **to** $16$

Figure 4 shows the simulation results by varying the switching overhead $Esw$ from 0 to $1mJ$. When $Esw$ is small, the sleep threshold is negligible, and hence the improvement of LA-Reallocation is small. However, when $Esw$ rises, the shutdown penalty for LA-DVS also increase, which makes LA-DVS even consume more energy than the Pure-DVS ($Esw > 0.5mJ$). On the other hand, since LA-Reallocation reduces the times of core switching, it is less affected by the increased $Esw$, and therefore yields more energy saving than LA-DVS.

Figure 5 shows the simulation results by varying the average ratio of actual execution time to WCET. The average total utilization is set to 0.3, which is close to the critical scale factor. The increased ratio shortens idle intervals, and hence it becomes more difficult for LA-DVS to take advantage of the idle intervals for core sleep. LA-Reallocation is less affected by the increased ratio. Figure 6 shows the simulation results by varying the core number from 2 to 16. While Pure-DVS results in a rapid increase in energy consumption with the growth of core number, the energy consumption under LA-Reallocation remains steady.

## 7. Conclusions

This paper has presented a run-time task reallocation scheme that improves the energy efficiency of leakage-aware DVS on multicore processors. Task reallocation is employed to dynamically merge the fragmentary idle intervals and to reduce the times of failure in turning relevant cores into sleep. Simulations have been conducted and the results demonstrate the effectiveness of the proposed scheme.

## 8. Acknowledgements

This work was partially supported by the National Natural Science Foundation of China under Grant No. 60703101 and No. 60903153, Zhejiang Provincial Natural Science Foundation of China under Grant No. Y108685, and the Fundamental Research Funds for the Central Universities (DUT).